\documentclass[12pt, preprint]{aastex}

\slugcomment{Not to appear in Nonlearned J., 45.}

\shorttitle{Dearth of low luminosity GC LMXBs}
\shortauthors{Fabbiano et al.}

\begin{document}


\title{The Dearth of low-luminosity Globular Cluster LMXBs in NGC 3379}


\author{G. Fabbiano}
\affil{Harvard-Smithsonian Center for Astrophysics, 60 Garden St., Cambridge, MA 02138}
\email{pepi@cfa.harvard.edu}

\author{N. J. Brassington, A. Zezas}
\affil{Harvard-Smithsonian Center for Astrophysics, 60 Garden St., Cambridge, MA 02138}

\author{S. Zepf}
\affil{Department of Physics and Astronomy, Michigan State University, East Lansing, MI 48824-2320}

\author{L. Angelini}
\affil{Laboratory for X-ray Astrophysics, NASA Goddard Space Flight Center, Greenbelt, MD 20771}

\author{R. L. Davies}
\affil{Sub-Department of Astrophysics, University of Oxford, Oxford OX1 3RH, UK}

\author{J. Gallagher}
\affil{Department of Astronomy, University of Wisconsin, Madison, WI 53706-1582}

\author{V. Kalogera}
\affil{Northwestern University, Department of Physics and Astronomy, Evanston, IL 60208}

\author{D.-W. Kim}
\affil{Harvard-Smithsonian Center for Astrophysics, 60 Garden St., Cambridge, MA 02138}

\author{ A. R. King}
\affil{Theoretical Astrophysics Group, University of Leicester, Leicester LE1 7RH, UK}

\author{A. Kundu}
\affil{Department of Physics and Astronomy, Michigan State University, East Lansing, MI 48824-2320}

\author{S. Pellegrini}
\affil{Dipartimento di Astronomia, Universita’ di Bologna, Via Ranzani 1, 40127 Bologna, Italy}

\and

\author{G. Trinchieri}
\affil{INAF-Osservatorio Astronomico di Brera, Via Brera 28, 20212 Milano, Italy}



\begin{abstract}
Our campaign of deep monitoring observations with {\it Chandra} of the nearby elliptical galaxy NGC 3379 has lead to the detection of nine globular cluster (GC) and 53 field low mass X-ray binaries (LMXBs) in the joint {\it Hubble}/{\it Chandra} field of view of this galaxy. Comparing these populations, we find a highly significant lack of GC LMXBs at the low (0.3-8~keV) X-ray luminosities (in the $\sim 10^{36}$ to $\sim 4\times10^{37}$ erg s$^{-1}$ range) probed with our observations. This result conflicts with the proposition that all LMXBs are formed in GCs. This lack of low-luminosity sources in GCs is consistent with continuous LMXB formation due to stellar interactions and with the transition from persistent to transient LMXBs. The observed cut-off X-ray luminosity favors a predominance of LMXBs with main-sequence donors instead of ultra-compact binaries with white-dwarf donors; ultra-compacts could contribute significantly only if their disks are not affected by X-ray irradiation. Our results suggest that current theories of magnetic stellar wind braking may work rather better for the unevolved companions of GC LMXBs than for field LMXBs and cataclysmic variables in the Galaxy, where these companions may be somewhat evolved.
\end{abstract}


\keywords{globular clusters: general --- X-rays: binaries --- galaxies: NGC 3379}



\section{Introduction}

Since their discovery in the Milky Way (see Giacconi 1974), the origin and evolution of Low-mass X-ray binaries (LMXBs) has been the subject of much discussion. LMXBs are found in both the stellar field and globular clusters (GCs). Their 
incidence per unit stellar mass is much higher in GCs than in the field, requiring a special formation mechanism, presumably dynamical (Clark 1975; Katz 1975). The high efficiency of these dynamical processes  led to the suggestion that all LMXBs may form in GCs and then disperse in the field (see e.g., Grindlay 1984; Grindlay \& Hertz 1985); however, some primordial field binaries are also expected to evolve into LMXBs, suggesting that there may be two populations of these sources, with distinct formation and evolutionary histories (see review by Verbunt \& van den Heuvel 1995). 

The discovery of X-ray source populations in early-type galaxies with {\it Chandra} has provided a wider and more diverse observational basis for the study of LMXBs, their association with GCs and the role that GCs may play in LMXB formation (see review Fabbiano 2006). However, until recently, only the most luminous extra-Galactic LMXBs with ($\sim$0.3-8~keV) $L_X \geq$ a few $10^{37}$erg s$^{-1}$ have been observed, and therefore the study of the GC-LMXB association has been limited to systems with X-ray luminosities in the upper range of Galactic LMXB luminosities.

With our deep 337 ks {\it Chandra} ACIS-S3 observations of the  unperturbed elliptical galaxy NGC 3379 in the nearby poor group Leo (D$\sim$11 Mpc), we can now pursue this study in a  luminosity range more typical of the well-studied Galactic LMXBs. In NGC 3379, The GC-LMXB association has been previously studied by Kundu, Maccarone \& Zepf (2007, hereafter KMZ), using the first shorter {\it Chandra} observation of this galaxy, which has a typical source detection threshold of $\sim 1-2 \times 10^{37}$ erg s$^{-1}$  (KMZ). Our data set is $\sim$10 times deeper (337 ks), allowing the detection of LMXBs at luminosities of $\sim 10^{36}$ erg s$^{-1}$. 

\section{The Lack of Low-Luminosity GC LMXBs}

We refer to our companion paper for details on the data, analysis and X-ray source catalog (Brassington et al 2007a, hereafter B07). The data set includes a first archival $\sim$30 ks pointing in 2001 Feb. (obsid 1587, used by KMZ), and four longer exposures distributed between 2006 Jan. and 2007 Jan. B07 find 9 GC-LMXB matches with the {\it Hubble} WFPC2 uniform set of optical GC detections of Kundu \& Withmore (2001). These GC-LMXBs are distributed throughout the entire luminosity range to which we are sensitive (see fig. 9 of B07), but are relatively scarce in the (1-10)$\times10^{37}$ erg s$^{-1}$ range, where instead, the overall number of sources sharply increases. 

We have explored the difference in the luminosity distributions of GC and field LMXBs suggested by the B07 results in three different ways, using only sources from the joint {\it Hubble}/{\it Chandra} field of view. We compared: 1) the number of sources detected by KMZ with those in our 10-fold deeper data set (Table 1); 2) the observed cumulative luminosity functions (source numbers) of field and GC LMXBs, which suffer from comparable incompleteness biases; and 3) the cumulative luminosity distributions (as function of luminosity), including the limit on the low luminosity emission of GC LMXBs, derived from a stacking experiment, which overcomes detection incompleteness.

It is clear from Table 1 that there is a relative lack of GC-LMXB associations at the lower LMXB luminosities, since the fraction of GCs associated with an LMXB does not change significantly as a result of the deeper detection thresholds, while the number of detected LMXBs increases by a factor of 2.4. If the number of GC-LMXB associations increased linearly with the number of LMXBs at the lower luminosities, we would expect to detect 17 GC-LMXB associations, instead of the 9 we detect; based on Poisson statistics, detecting 9 sources has a low chance probability of 1.2\%. Of these 9 sources, 6 are associated with red GCs, and 3 are found in blue GCs (using V−I = 1.0 as a boundary, from Table 11 of B07); although we only have small numbers of GC sources, the predominance of red GCs is consistent with previous studies of GC LMXB populations (e.g., KMZ, Kim E. et al 2006). 

Fig. 1 shows the observed cumulative X-ray luminosity functions (XLFs) of field and GC LMXBs. These XLFs are not corrected for incompleteness at the low luminosities, which becomes important below $\sim8\times10^{36}$ erg s$^{-1}$. However, the two samples of sources were extracted from the same field, therefore both XLFs suffer from comparable detection biases, and can be compared directly. It is evident that the GC LMXB XLF lacks low luminosity sources, compared to that of field LMXBs. A KS test excludes that the two distributions may be derived from the same parent population at 99.82\% confidence. This is a conservative estimate, because the field-LMXB XLF contains a larger number of sources from the circum-nuclear regions, which in principle should be more affected by detection incompleteness because of the higher background from the hot ISM and source confusion. Therefore the statistical significance of the difference can only increase if all biases were removed. 

The dearth of low luminosity GC LMXBs is reinforced by the results of a stacking experiment, on the GCs with undetected X-ray counterpart. We created 56 source regions, centered on the location of the GCs from Kundu \& Withmore (2001; 14 of these 70 GCs were not included, as they either had confirmed X-ray counterparts, or were too close to multiple X-ray sources for reliable photometry, see B07). The radii of these 56 regions were set at the 1.5 keV 95\% encircled energy radius, with a minimum of 3" near the aim point. Background counts were extracted from annuli concentric with source regions, with inner and outer radii of two and five times the source radius respectively, excluding areas with detected  B07 sources. We obtain 1229 cumulative counts from the source regions, and 1239.3 background counts (normalized by the ratio of extraction areas).
Fig.~2 shows the histograms of the source counts (lower panel) and background subtracted net counts (upper panel) extracted from each stacking regions; the median value of the net count distribution is -0.08, showing that there are no biases in the determination of the background counts.

Following the same Bayesian approach used in B07, which takes into account the Poisson nature of the probability distribution of the source and background counts, as well as the effective area at the position of the source (van Dyk et al. 2001; Park et al 2006), we find upper bounds on the intensity of a `stacked' source of 28.8 and 101.1 counts at 68\% and 99.7\% confidence, respectively (in both cases the mode is 0).  Dividing by the number of GCs included in the experiment, source flux and luminosity were calculated in the 0.3-8.0 keV band, with an energy conversion factor (ECF) corresponding to an assumed power law spectral shape, with $\Gamma=1.7$ and Galactic $N_H=2.78\times10^{20 }$ cm$^{-2}$, where the ECF also accounts for the temporal and spatial variations of the ACIS CCD quantum efficiency\footnote{See \url{http://asc.harvard.edu/cal/Acis/Cal\_prods/qeDeg for the low energy QE degradation}}  and the vignetting effect (Kim et al 2004). We obtain average upper confidence bounds on the X-ray luminosity of a non-detected GC $< 1.2\times10^{35}$ erg s$^{-1}$ (68\%) and $< 5.4\times10^{35}$ erg s$^{-1}$ (99.7\%); these limits are well below the distribution of luminosities of detected LMXBs.  

Fig.~3 shows the cumulative luminosity distributions of field and GC LMXBs; the stacking upper limit excludes at high statistical significance that incompleteness is responsible for the lack of GC sources at the low luminosities and shows that these sources are really absent down to very low limits. Survival analysis tests (from ASURV; Lavalley, Isobe \& Feigelson 1992) on these distributions show that the GC and field populations are different at the 100\% confidence level (both Logrank and Peto-Prentice tests). 

\section{Variability Properties of GC LMXBs}

B07 identifies seven variable GC sources (out of nine); light-curves of all these sources can be found in B07. Of the three most luminous GC LMXBs, two are highly variable between observations: sources 41 of B07 (with average 0.3-8.0 keV $L_X=2\times10^{38}$ erg s$^{-1}$) and 42 ($L_X=5\times10^{38}$ erg s$^{-1}$); the third (source 67, $L_X=3\times10^{38}$ erg s$^{-1}$) does not vary. No short-term variability, within each observation, was detected in B07 in these sources. 

Five of the GC LMXB sources with $L_X < 10^{38}$ erg s$^{-1}$ vary. Three of these sources are detected only in a single observation (sources 50, 61, 79; B07); the most luminous, source 50, is a highly significant transient candidate, with a detected luminosity of $6.3\times10^{37}$ erg s$^{-1}$, a factor of $\sim$20 times larger than the non-detected threshold luminosity. Sources 61 and 79 are fainter: 61 has a peak detection luminosity of $1.6\times10^{37}$ erg s$^{-1}$ and a ratio of peak to non-detected of 7.5, making it a possible transient candidate (see B07); 79 is a highly significant transient candidate (ratio of peak to non-detected luminosity of 22) with a peak luminosity of $1.3\times10^{37}$ erg s$^{-1}$. Source 50 is detected in the first observation after a 5-year observation hiatus, so that we could be catching the end of a few-years on-state ($< 5$ yr); source 61 is detected only in the first observation, so the duration of its on-state is unconstrained; source 79'      s on-state instead can be constrained by our sampling to be $< 9$ months (see also Brassington et al 2007b, in preparation).
 
\section{Discussion}
Our results  strongly suggest  a dearth of GC-LMXB associations at luminosities lower than $\sim4\times10^{37}$ erg s$^{-1}$ in the 0.3-8~keV band. Above this luminosity, KMZ estimate a 5\% fraction of GCs with LMXBs, similar to that observed in other elliptical and S0 galaxies for a comparable luminosity threshold (KMZ, see also Fabbiano 2006 and references therein). The comparison of the XLFs of field and GC LMXBs of E and S0 galaxies (Kim, E. et al 2006) shows that the two track each other above $\sim4\times10^{37}$ erg s$^{-1}$. In NGC 3379, for the first time we can probe the the low-luminosity field and GC XLFs of an elliptical galaxy and find that they differ, suggesting a decrease in the number of GCs associated with LMXBs at luminosities $L_X < \sim 4\times10^{37}$ erg s$^{-1}$. 
A similar effect was suggested by the comparison of GC and Field XLFs in NGC3115, where source detection extends below $10^{37}$ erg s$^{-1}$ (see fig. 6 of KMZ). The study of GC-LMXB populations in M31 and the Milky Way (although distance uncertainties may play a role in the latter) are also consistent with this emerging picture (see figures 6 and 7 of Voss \& Gilfanov 2007); in both cases,  the differential XLFs of field and GC LMXBs differ at the lower luminosities, with the GC LMXB XLF showing a sharp downturn at luminosities lower than $\sim 5\times10^{36}$ erg s$^{-1}$. Our deep look at NGC 3379 suggests that this behavior may be a general feature of LMXB populations. 

One cause for this effect, which we can discount, could be the presence of multiple LMXBs in the most luminous GCs that might 'remove' sources from the fainter portion of the XLF. However, source variability such as we detect in GC LMXBs (see Section 3 and B07) suggests that multiple LMXB systems are unlikely to be found in the most luminous GCs sources, a conclusion in agreement with the independent evidence of Sivakoff et al (2007).

A possible explanation for the difference in GC and field XLFs could be a relative excess of high luminosity LMXBs in GCs, because of the expected overwhelming presence of transients at the high luminosities in the old stellar field population of elliptical galaxies (Piro \& Bildsten 2002; King 2002). GC binaries with either main sequence (MS),  red giant, or white dwarf donors, instead, can be bright, persistent X-ray sources, because they form predominantly by stellar interactions (Clark 1975; Katz 1975; Ivanova et al. 2007) and so escape the age constraints of primordial field binaries. The only ways of making transients in GCs are rather unlikely, including either the capture of a giant star - rare, because of their short lifetimes, or a BH+main-sequence binary evolving to BH+giant - also rare, because the nuclear evolution has to beat angular momentum losses. Short-period transients in the field have nuclear-evolved companions because of a complex previous evolution that does not apply to GC sources made by dynamical stellar interactions.

Even black hole (BH) sources in GCs could be persistent (Kalogera, King \& Rasio 2004). High luminosity GC LMXBs may be BH binaries, given their luminosities near or above the Eddington limit of an accreting NS and their widespread variability (this paper and Maccarone et al 2007); the most luminous GC source we find in NGC 3379 (in a red cluster, B07) has a luminosity of $5\times10^{38}$ erg s$^{-1}$ and is variable. This peak luminosity is consistent either with slightly (by a factor of $\sim 2$) super-Eddington luminosities for a NS accreting H-rich material or with Eddington luminosity for He-rich material.The detection of three {\em possible} transients in the NGC 3379 GC sources, with peak luminosity in the $10^{37}$ erg s$^{-1}$ range, may argue against mostly persistent sources, but we must remember that typical peak-to-quiescence flux ratios for transients are $\gtrsim$100, a range we cannot explore with our sensitivity (see \S3.).

Alternatively, we may have a real lack of low-luminosity GC sources because of the transition from persistent to transient X-ray sources, as discussed by Bildsten \& Deloye (2004, BD4 hereafter) in the context of their model of ultra-compact binaries (UCs) with NS accreting from white dwarfs. This conclusion is supported by the very low limits on the luminosity of undetected GC sources found in our stacking experiment (\S2.).
In the disk instability model (King et al 1997) this transition occurs when the mass transfer rate driven from the donor drops below a critical value. Since in persistent X-ray sources the mass transfer rate is thought to be directly connected to the X-ray luminosity, this transition would lead to a dearth of X-ray sources with luminosity lower than the one corresponding to the critical mass transfer rate. BD4 showed that persistent UCs must have a high-luminosity XLF consistent with the observed high luminosity XLFs of both GC and field LMXBs (see Kim \& Fabbiano 2004; Kim, E. et al 2006). However, quantitative consideration of this cut-off X-ray luminosity for UCs with He-rich donors leads to rather low cut-off values. Figure 3 in Deloye \& Bildsten (2003) indicates a value of $\sim 5\times10^{36}$ erg s$^{-1}$ for {\em non-irradiated} accretion disks, and the cut-off would occur at significantly lower values for irradiated disks (see, for example, King, Kolb, \& Burderi 1996). 

There is overwhelming evidence that Galactic LMXB accretion discs are significantly irradiated. van Paradijs \& McClintock (1994) showed that irradiation fixes their absolute visual magnitudes, and van Paradijs (1996) and King, Kolb \& Burderi (1996) that irradiation determines whether these systems are transient or not. There is of course no reason why this conclusion should change for the GC LMXBs in NGC 3379. Accordingly, we conclude that the observed cut-off luminosity identified here is not consistent with the suggestion that the majority of the GC LMXBs are UCs. We note that the recent self-consistent population simulations with dynamical interactions presented by Ivanova et al (2007) also suggest that UCs may not be the dominant LMXB population in GCs, contrary to BD4's assertions. Formation rates of UCs are found to be comparable or lower than those of LMXBs with MS donors (see their figure 3), whereas UC lifetimes are found to be $\sim10^7 - 10^8$\,yr (also C.\ Deloye, private communication), lower or at best comparable to the lifetime of the persistent phase for LMXBs with MS donors.  Moreover, Ivanova et al. (2007) do not find a significant dependence of the number and evolution of ultra-compacts on the metallicity of the parent GC, contrary to the prevalent association of LMXBs with red metal-rich GCs in elliptical galaxies (e.g., KMZ; Kim E. et al. 2006).

In view of the rather high observed cut-off luminosity of the GC LMXBs and the results of Ivanova et al.\ (2007) we conclude that the GC sources in NGC~3379 are not consistent with UCs being the predominant population. Of the other two donor types, red giants and MS, red giants are also disfavored because they are expected to be transient through most of their lifetime (King et al.\ 1997); consequently they could not contribute to the formation of a X-ray luminosity cut-off. The simulations by Ivanova et al. (2007) also indicate that their numbers are low, taking into account their formation rates, short lifetimes truncated by their large cross sections and frequent stellar interactions, and their transient duty cycles. LMXBs with H-rich MS donors remain as the only possibility for a dominant, persistent X-ray source population in GCs that would produce the bright systems we observe. For irradiated H-rich accretion disks MS donors are expected to transition from persistent to transient behavior at $\sim 3\times10^{37}$\,erg\,s$^{-1}$ (King, Kolb, \& Burderi 1996; Stehle, Kolb, \& Ritter 1997); this transition luminosity would also be consistent with the observed peak luminosities of our transient candidates.

Our results show that any persistent GC LMXBs with MS companion is likely to have $L_X > 4 \times 10^{37}$ erg s$^{-1}$. A qualitatively plausible interpretation is that mass transfer in these bright systems is driven by magnetic braking, in the standard form for Pop II systems (Stehle, Kolb \& Ritter, 1997); instead in sources fainter than $\sim 4 \times 10^{37}$erg~s$^{-1}$ mass transfer would be driven  by gravitational radiation. It is noticeable that this simple interpretation does not work so well for field LMXBs or cataclysmic variables (CVs) in the Galaxy, as magnetic braking does not seem to produce such uniformity. This may relate to the fact that the companion stars in GC LMXBs are of necessity completely unevolved main sequence stars, since they have been dynamically captured. By contrast, the companion stars in field LMXBs and CVs have probably had a previous complex interaction with their neutron star/white dwarf primaries, and may well be somewhat evolved despite their current low masses.

There is an important immediate implication of our results for the understanding of LMXB evolution in galaxies: the difference between field and GC LMXB luminosity functions at $L_X < 4\times10^{37}$ erg s$^{-1}$ excludes a single formation mechanism in GCs for all LMXBs, resolving a long-standing controversy in LMXB formation and evolution (e.g., Grindlay 1984; Grindlay \& Hertz 1985; review by Verbunt \& van den Heuvel 1995). The proposition that all field LMXBs were formed in GCs is incompatible with the significant difference between their luminosity function and that of GC LMXBs. This conclusion is in agreement with the observations of the Sculptor dwarf spheroidal galaxy (Maccarone et al 2005), suggesting that the binary properties of field and GC LMXBs might be different (see KMZ), and with less direct early suggestions, based on LMXB and GC population statistics in elliptical galaxies (Irwin 2005; Juett 2005).

\section{Conclusions}
Our campaign of deep monitoring observations of the nearby elliptical galaxy NGC 3379 with {\it Chandra} ACIS-S3 has led to the detection of  nine  GC LMXBs in the field studied optically with {\it Hubble} WFPC2,  seven of which were previously detected in a much shorter {\it Chandra} exposure (1/10 of the exposure time; KMZ). The comparison of GC and field LMXB statistics in the joint {\it Hubble}-{\it Chandra} field of view demonstrates a relative lack of GC LMXB at luminosities below $\sim 4\times10^{37}$ erg s$^{-1}$; field and GC LMXBs instead are know to have closely matching XLFs above this luminosity (Kim E. et al 2006; KMZ). The dearth of low-luminosity GC LMXBs in NGC 3379 is consistent with a similar suggested behavior in NGC 3115 (KMZ), and with the XLFs of field and GC LMXBs in the Milky Way and M31 (Voss \& Gilfanov 2007). These differences between low-luminosity field and GC XLFs falsify suggested theories that {\it all} LMXBs may have been originated in GCs. 

The luminosity-dependent differences of field and GC XLFs cannot be explained by high luminosity GCs containing multiple LMXBs, because we find clear evidence of source variability for seven out of the nine sources invalidating this hypothesis. Persistent behavior of high luminosity GC sources, compared with transient field sources of similar high luminosity may explain the discrepancy as excess of luminous GC LMXBs. However, the detection of three candidate transient source (with peak luminosity greater than $10^{37}$ erg s$^{-1}$) in the GC LMXB population of NGC 3379 may not support this explanation. 

The lack of low-luminosity sources in GCs is consistent with the prediction of Bildsten and Deloye (BD4) based on the transition of sources from persistent to transients due to the thermal disk instability. However, the value of the observed XLF cut-off is not  consistent with their suggestion that GC LMXBs are dominated by ultra-compact binaries, but instead favors LMXBs with H-rich MS donors. The $\sim 4\times10^{37}$ erg s$^{-1}$ luminosity cut-off is also consistent with current theories of magnetic stellar wind braking, suggesting that this effect may work rather better for the unevolved companions of GC LMXBs than for field LMXBs and cataclysmic variables in the Galaxy, where these companions may be somewhat evolved.

While our results firmly establish a dearth of GC sources in NGC3379 at low luminosity, the accurate luminosity (and uncertainty) of the GC XLF  cut-off will need the formal analysis of the NGC3379 LMXB luminosity function (Kim et al in preparation). The forthcoming analysis of the very deep {\it Chandra} observations of NGC4278, a GC-rich galaxy (see Kim, D.-W. et al 2006), will provide in the near future stronger constraints on the potential 'universality' and value of the GC LMXB XLF cut-off luminosity.



\acknowledgments
We thank Chris Deloye and Natalia Ivanova for very very useful discussions.  
The data analysis was supported by the CXC CIAO software and CALDB. We have used the NASA NED and ADS facilities, and have extracted archival data from the {\it Chandra} archives. This work was supported by the {\it Chandra}
 GO grant G06-7079A (PI: Fabbiano) and sub-contract G06-7079B (PI: Kalogera). We acknowledge partial support from NASA contract NAS8-39073 (CXC); A. Zezas acknowledges support from NASA LTSA grant NAG5-13056; S. Pellegrini acknowledges partial financial support from the Italian Space Agency ASI  (Agenzia Spaziale Italiana) through grant ASI-INAF I/023/05/0. 



{\it Facilities:}  \facility{CXO (ACIS)}.

\clearpage




\begin{figure}
\plotone{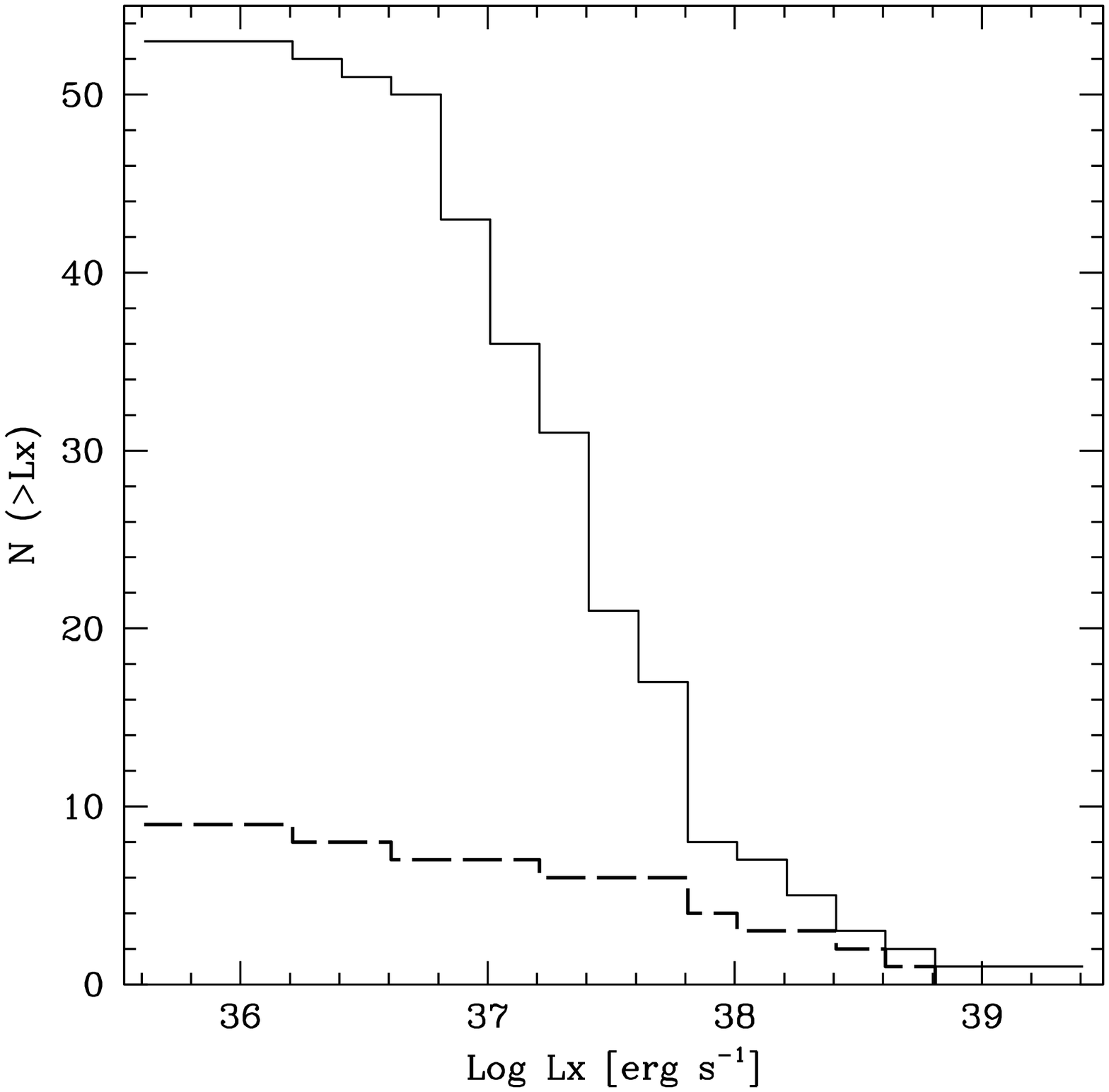}
\caption{Observed cumulative XLFs of field (continuous line) and GC (dashed) LMXBs; both distributions are from the joint Hubble/Chandra field.}
\end{figure}

\begin{figure}
\plotone{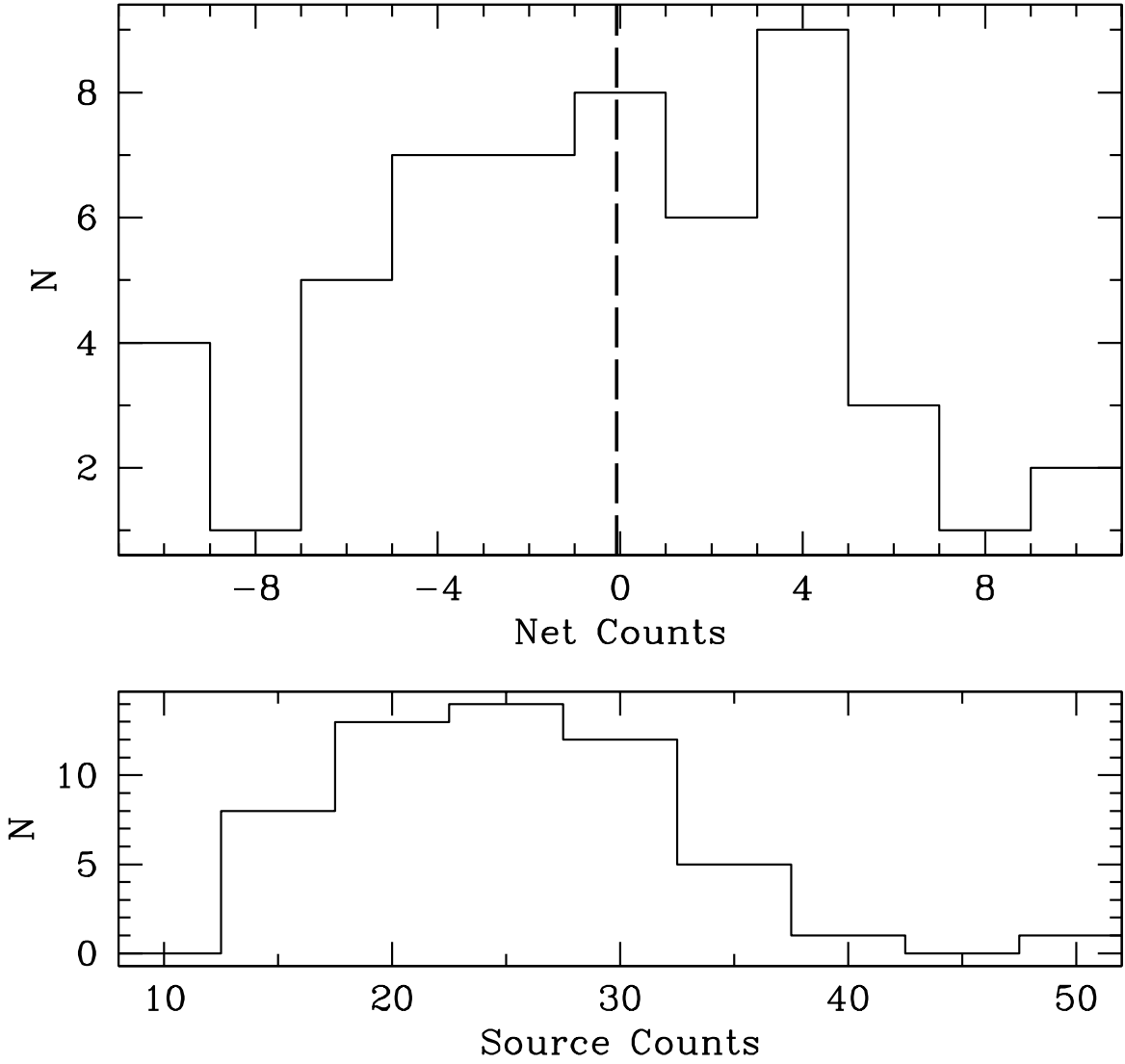}
\caption{Histograms of source counts (lower panel) and background-subtracted net counts (upper panel) for the sources included in the stacking. The vertical line in the upper panel is the median of the distribution.}
\end{figure}

\begin{figure}
\plotone{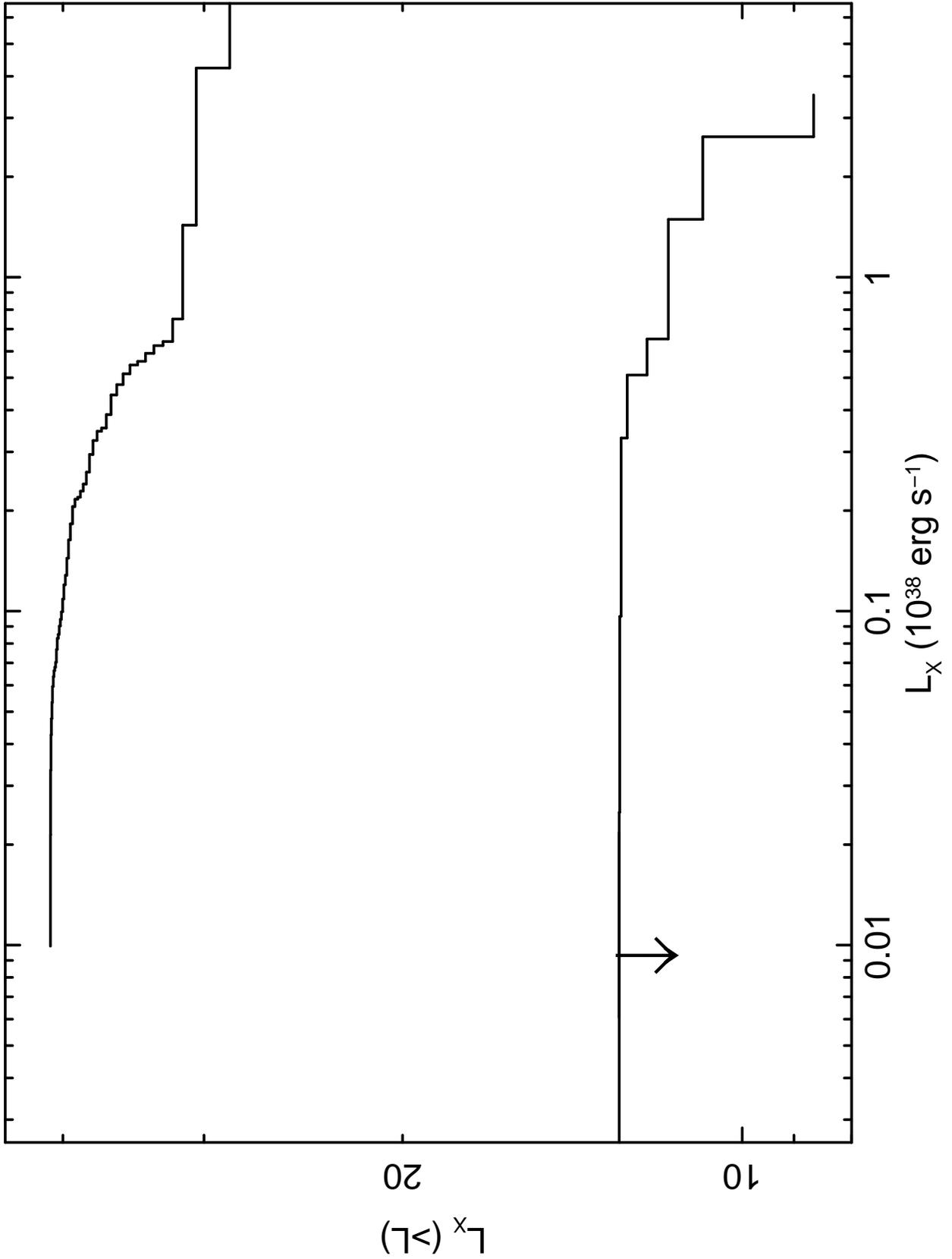}
\caption{Cumulative luminosity distributions of detected field (black) and GC (red) LMXBs; both distributions are from the joint Hubble/Chandra field. The last bin with thearrow represents the contribution of undetected GC LMXB from our stacking experiment.}
\end{figure}

\clearpage

\begin{table}
\begin{center}
\caption{LMXB-GC Associations in NGC 3379}
\begin{tabular}{cccccccc}
\tableline\tableline
Observations\tablenotemark{a} & $T_{exp.}$(ks) & GC\tablenotemark{b} & LMXB\tablenotemark{c} & GC-LMXB & $f_{GC}$ & $f_{LMXB}$  & Ref.\\
\tableline
1587 & 31 & 61 & 26 & 7 & 12\% & 27\% &	KMZ\\
1587, 7073, 7074, 7075,7076	&337 	&70	&62	&9 &	13\%	&15\%	&B07\\
\tableline
\end{tabular}
\tablenotetext{a}{These observations were all performed with {\it Chandra} ACIS-S3; 1587 was obtained in 2001, the other four observations were all taken at few months intervals between 2006 Jan. and 2007 Jan. (see B07).}
\tablenotetext{b}{Number of GCs within the WFPC2 observation of NGC 3379 considered for the LMXB search; no GC was detected within 5'' of the nucleus (Kundu \& Withmore 2001, with color cut of 0.5-1.5 in V-I). Note that the number of GCs in KMZ is slightly smaller because of their restrictive color cut (0.8-1.4 in V-I, see KMZ).}
\tablenotetext{c}{Number of all the LMXB detected in the WFPC2 area, including 5 sources detected within 5'' of the nucleus where no GC counterpart was detected. Since LMXBs may be confused with {\it Chandra} in this inner area, $f_{LMXB}$ could be somewhat overestimated. Confusion would affect more the deeper cumulative observation of B07, strengthening our conclusion of a lack of GC sources at the lower luminosities}
\end{center}
\end{table}

\end{document}